\documentclass[10pt,aps,pre,twocolumn,notitlepage,superscriptaddress,preprintnumbers]{revtex4-1}
\usepackage[utf8]{inputenc}

\usepackage{amsmath,amssymb,amsfonts}
\usepackage{bm}
\usepackage{graphicx,color}
\usepackage{verbatim}
\usepackage{float}
\usepackage[caption = false]{subfig}
\usepackage{dcolumn}
\usepackage{natbib}
\usepackage{hyperref}
\usepackage{tikz}
\usetikzlibrary{hobby}
\usepackage{appendix}
\usepackage{ulem}

\newcommand{\nodelink}{\mathbin{\rotatebox[origin=c]{180}{$\multimap$}}}

\newcommand*\nodeLinkNode{\includegraphics[scale=0.3]{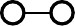}}

\begin{document}
	
	\title{Second to first order phase transition; coevolutionary versus structural balance }
	\author{M. Ghanbarzadeh Noudehi}
	\email{mgh.setareh@gmail.com}
	\affiliation{Department of Physics, Institute for Advanced Studies in Basic Sciences, 45195-1159, Zanjan, Iran}
	\author{A. Kargaran}
	\affiliation{Department of Physics, Shahid Beheshti University, Evin, Tehran 19839, Iran}
	\author{N. Azimi-Tafreshi}
	\affiliation{Department of Physics, Institute for Advanced Studies in Basic Sciences, 45195-1159, Zanjan, Iran}
	\author{G. R. Jafari}
	\email{g\_jafari@sbu.ac.ir}
	\affiliation{Department of Physics, Shahid Beheshti University, Evin, Tehran 19839, Iran}
	\affiliation{Institute of Information Technology and Data Science, Irkutsk National Research Technical University, 83, Lermontova St., 664074 Irkutsk, Russia}
	
\date{\today}
	
\begin{abstract}
In social networks, the balance theory has been studied by considering either the triple interactions between the links (structural balance) or the triple interaction of nodes and links (coevolutionary balance). In the structural balance theory, the links are not independent from each other, implying a global effect of this term and it leads to a discontinuous phase transition in the
system’s balanced states as a function of temperature. However, in the coevolutionary balance the links only connect two local nodes and a continuous phase transition emerges. In this paper, we consider a combination of both in order to understand which of these types of interactions will identify the stability of the network. We are interested to see how adjusting the robustness of each term versus the other might affect the system to reach a balanced state. We use statistical mechanics methods and the mean field theory and also the Monte-Carlo numerical simulations to investigate the behaviour of the order parameters and the total energy of the system. We find the phase diagram of the system which demonstrates the competition of these two terms at different ratios against each other and different temperatures. The system shows a tricritical point above which the phase transition switches from continuous to discrete. Also the superiority of the local perspective is observed at low temperatures and the global view will be the dominant term in determining the stability of the system at higher temperatures.

\end{abstract}
\maketitle

\section{Introduction}\label{Introduction}

Signed networks with two types of positive and negative interactions have been used to model dynamical processes in a wide range of scientific fields including social science \cite{altafini,belaza1,samin}, politics and international relationships \cite{hart,galam,schoors,thurner2}, biology \cite{saminbio,yensheng,abbas,zahra}and ecology \cite{saiz}. These networks exhibit the concept of structural balance which was first proposed by Heider  in social science based on triple relations \cite{heider1, heider2}. Heider defined a triple as balanced if the product of its link signs is positive, otherwise it is unbalanced. Later, Cartwright and Harary developed the mathematics of balance theory \cite{Cartwright}. They showed that a complete network is structurally balanced if either all the link signs are positive, or the network is divided into two clusters such that the signs of links are positive within each cluster and links between the clusters have negative signs.

The social tension of a system is measured by summing up all triple interactions in the network which can be defined as the Hamiltonian of the system \cite{Cartwright}. Antal et al., studied the dynamics of structural balance in social networks such that each link changes its sign in a way to decrease tension \cite{antal1}. In \cite{fereshteh} a model is proposed that takes into account the temperature, as a measure of tension tolerance, and its effect on the dynamics of social networks is studied. Using the mean-field analysis, they showed that a first-order phase transition emerges in the system’s balanced states as a function of temperature.

Further studies have been done considering the individuals' attitude toward a specific issue. In this case, a state is assigned to each node of the network besides the state of their relationships \cite{holm}. The individuals may build up their idea according to the idea of their friends \cite{sood, Castellano}, or they may update their relations in a way to friendly connect to people with the same idea of their own \cite{McPherson, Soderberg}. If we consider a unit of three objects (two nodes and one link), the agreement (disagreement) in the state of the nodes while their relation is friendship (enmity) will result in a balanced unit. In this model, called the coevolutionary balance model, we can put a sum over all such units in the network to obtain the social tension, which is described mathematically as the Hamiltonian of the system \cite{holm,Yan}. Despite the Heider balance model, this model represents a continuous phase transition in the energy of the system \cite{meghdad1,amir}.

In later studies comparisons have been made between these two structural and coevolutionary balance models in order to investigate the role of each balance view \cite{groski,thurner1,thurner3,meghdad2,amir}. The investigations showed the coevolutionary term, considering node-link-node interactions, supports local benefits for people in the network as its concern is reducing the tension in each person's neighboring. While the structural balance, with emphasizing on the triadic interactions, is in a direction to fulfill the global balance \cite{facchetti}. Researchers also studied the Network considering pairwise and higher order interactions in which more than two units are involved\cite{hakim}. People also used hypergraphs and simplicial complexes and showed that depending on the order of interaction, phase transition of the system will be continuous or discontinuous and this is similar to the compression of the results obtained in investigation of described balance models\cite{nature}. Here a question is raised that which of these approaches finally determines the balance state of the network.
Thurner et al. studied a model for the energy of a social network in which the Heider structural balance is evaluated in the presence of the coevolutionary balance \cite{thurner1}. They have measured the ratio of the differences between the number of balanced and unbalanced triangles to the total number of triangles through simulating on a small world network. Their results indicated that a society may fall into a fragmented or cohesive regime due to the dominance of each balance term. The structural balance contribution, extend the global view by constructing balanced triangles in the network and making the whole network to interact even if people might have different ideas and there is a cohesion in the society. However, the coevolutionary balance implies a local view and it shows that people are more likely to connect to those who have similar ideas in their own neighboring and the society will be shifted to a fragmented regime.

In this study, we consider the role of both balance terms in the system's evolution. We find the critical temperature and phase transitions of the system through analytical methods. Also we compare the results with the numerical simulations. In the analytical approach, we use the statistical mechanics methods, in equilibrium, to minimize the energy of the system, considering both triad and node-link-node interactions. The exponential random graph method  \cite{holland,besag,frank,strauss,wasserman,anderson,snijders1,robins,cranmer,snijders2} and the mean field approximation \cite{newman1,newman2,newman3} enable us to solve the problem for a fully connected network. We obtain the behavior of the correlation functions of the network as well as the total energy of the system versus temperature for different impacts on each balance term. Our study shows that the type of phase transitions changes from the continuous to the discontinuous by increasing the role of the Heider structural balance versus coevolutionary balance. We also present numerical simulation results on a small world network.

The paper is organized as follows. In the next section, we define the model and two-point correlation functions. In section III, we present the analytical method and, within the framework of mean-field theory, we find the average values of network variables. We obtain the fixed
points of the dynamics and analyze their stability. In section IV we present numerical simulations and compare the results with the analytical solutions. The paper is concluded in section V.

\section{Model}\label{model}

Let us consider a complete signed network with $n$ nodes. The state of each node $i$ is denoted by $s_i=\pm 1$ which shows the positive or negative opinion of each individual. Each link $(i,j )$, labeled with $\sigma_{ij}$, represents friendly (positive) and unfriendly (negative) relation between the nodes $i$ and $j$. Hence the number of total configurations is $2^n 2^{n(n-1)/2}$.

\begin{figure}[!t]
	\centering
	\includegraphics[width=\linewidth]{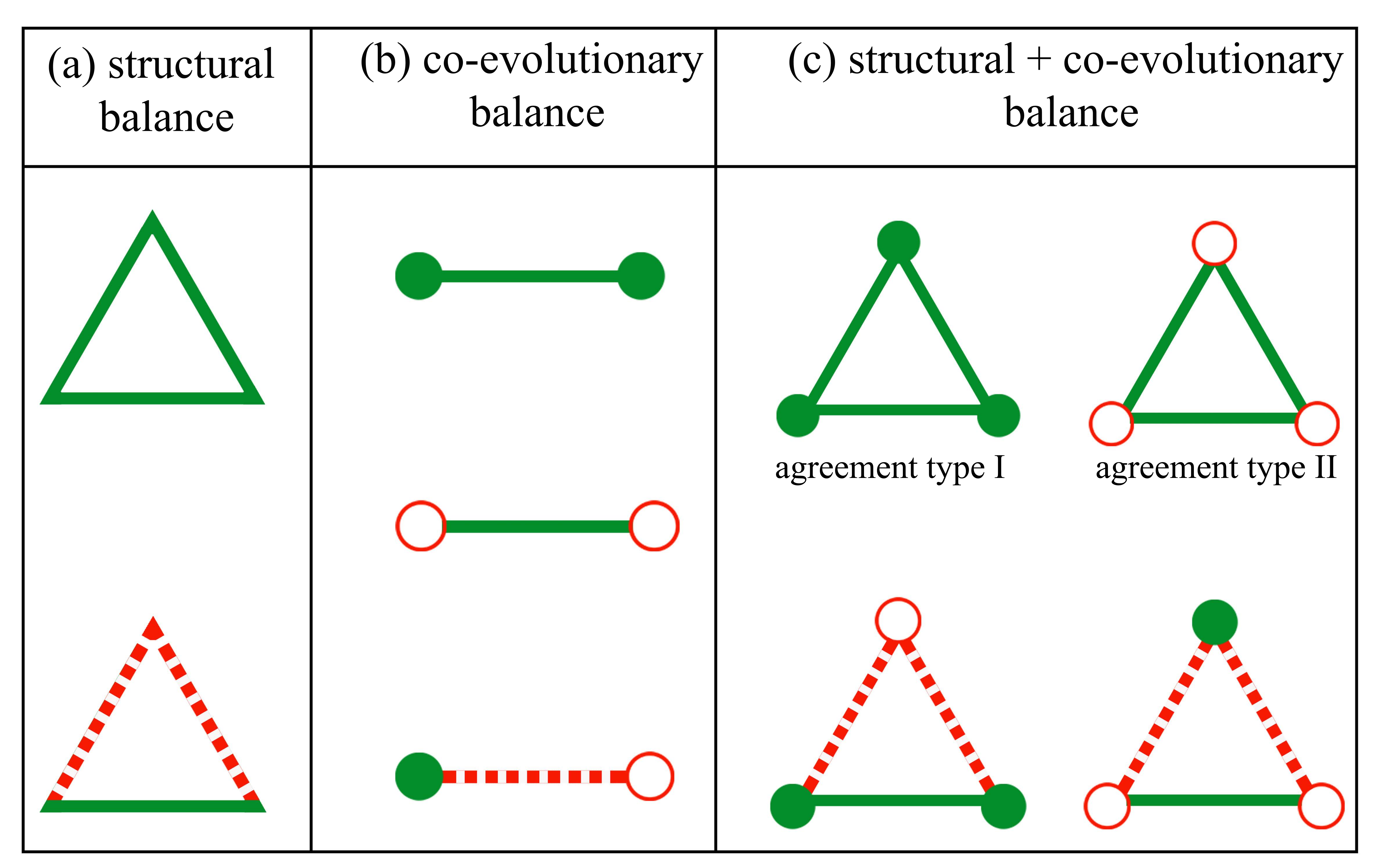}
	\caption{ Configurations of $(a)$ structural balance, $(b)$ coevolutionary balance and $(c)$  combination of the two structural and coevolutionary balance units.  The solid (green) and the dashed (red) lines show friendship and enmity connections respectively. Also the filled (green) and the empty (red) circles represent the positive and negative opinion of nodes respectively.
	}
	\label{fig:config}
	
\end{figure}

Considering the interaction of the links and also the interaction of a link with the end nodes, the total energy for a particular configuration $G$ is given by the following Hamiltonian,

\begin{equation}\label{eq1}
\mathcal{H}(G)=  -{\sum_{i<j} s_{i} \sigma_{ij} s_{j}} - g{\sum_{i<j<k} \sigma_{ij} \sigma_{jk} \sigma_{ki}} .
\end{equation}

The first term on the right-hand side of Eq.~(\ref{eq1}) shows the contribution of interactions of the links with their end nodes. In a society, this term describes the interactions between people's opinions and reaches a minimum when the coevolutionary balance occurs. Two friendly neighbors tend to share the same opinion while two enemy neighbors keep opposing opinions. This contribution of interactions implies how the dynamics of the system makes an individual converges to the opinion of his neighbors. The second term on the right-hand side of Eq.~(\ref{eq1}) represents the energy contribution due to the link interactions and implies the structural balance.
The factor $g$ controls the contribution of the structural balance with respect to the coevolutionary balance. The opinion formation and the sign of the links are in such a way that the total energy is minimized and the whole network reaches a balance state.
	
In the Heider theory, a triple is structurally balanced if the product of the link's signs is positive. It occurs if all the links are positive or if there exists only one positive link. Also, a unit consisting of a link and two end nodes is balanced in coevolutionary framework, if the product of the sign of the link and its end nodes is positive. In  combination of structural and coevolutionary balance, a triple is defined as balanced if: (1) all links are positive and the nodes either all positive (agreement type I) or all negative (agreement type II). (2)  there is only one positive link with two end nodes which have the same state.  Both end nodes are connected with negative links to the other node in the triple which has an opposite state. The balanced configurations are illustrated in Fig.~\ref{fig:config}.

In addition to the total energy, we can obtain the average value of some variables of the network. For instance, the average sate of the nodes and the links are defined respectively as $m\equiv\langle s_{i}\rangle$ and $p\equiv\langle\sigma_{ij}\rangle$. Also we define the following two-point correlations:

\begin{equation}\label{eq2}
q_{1}\equiv\langle s_{i}\sigma_{ij}\rangle,\quad q_{2}\equiv\langle\sigma_{ij}\sigma_{jk}\rangle,\quad q_{3}\equiv\langle s_{i}s_{j}.\rangle
\end{equation}

where $q_1, q_2$ and $q_3$ denote the node-link, the link-link and the node-node correlations respectively.

In the following we obtain the total energy and correlation functions of the model analytically and compare the results with the numerical simulations on a fully connected network.

\section{Analysis}\label{analysis}

Statistical physics and exponential random graph method give us the possibility to calculate the average value of a desired parameter $x$ on the all graph configurations using the Boltzman probability distribution $P(G)=\frac{e^{-\beta\mathcal{H}(G)}}{\mathcal{Z}}$, in which $\mathcal{Z}=\sum_{G} e^{-\beta\mathcal{H}(G)}$ is the partition function and $\beta=\frac{1}{T}$ is the inverse of temperature. Temperature $T$ shows the society tension and the fluctuation in opinion state or the type of relationship that people will make \cite{Bahr}. At low temperature, there is less tendency to change the state but when the temperature increases people show more interest in updating their states. The updates will affect the tension in the society by getting close to or far from the balance state of the system.
\begin{figure*}[!t]
	\begin{center}
		\scalebox{0.35}{\includegraphics[angle=0]{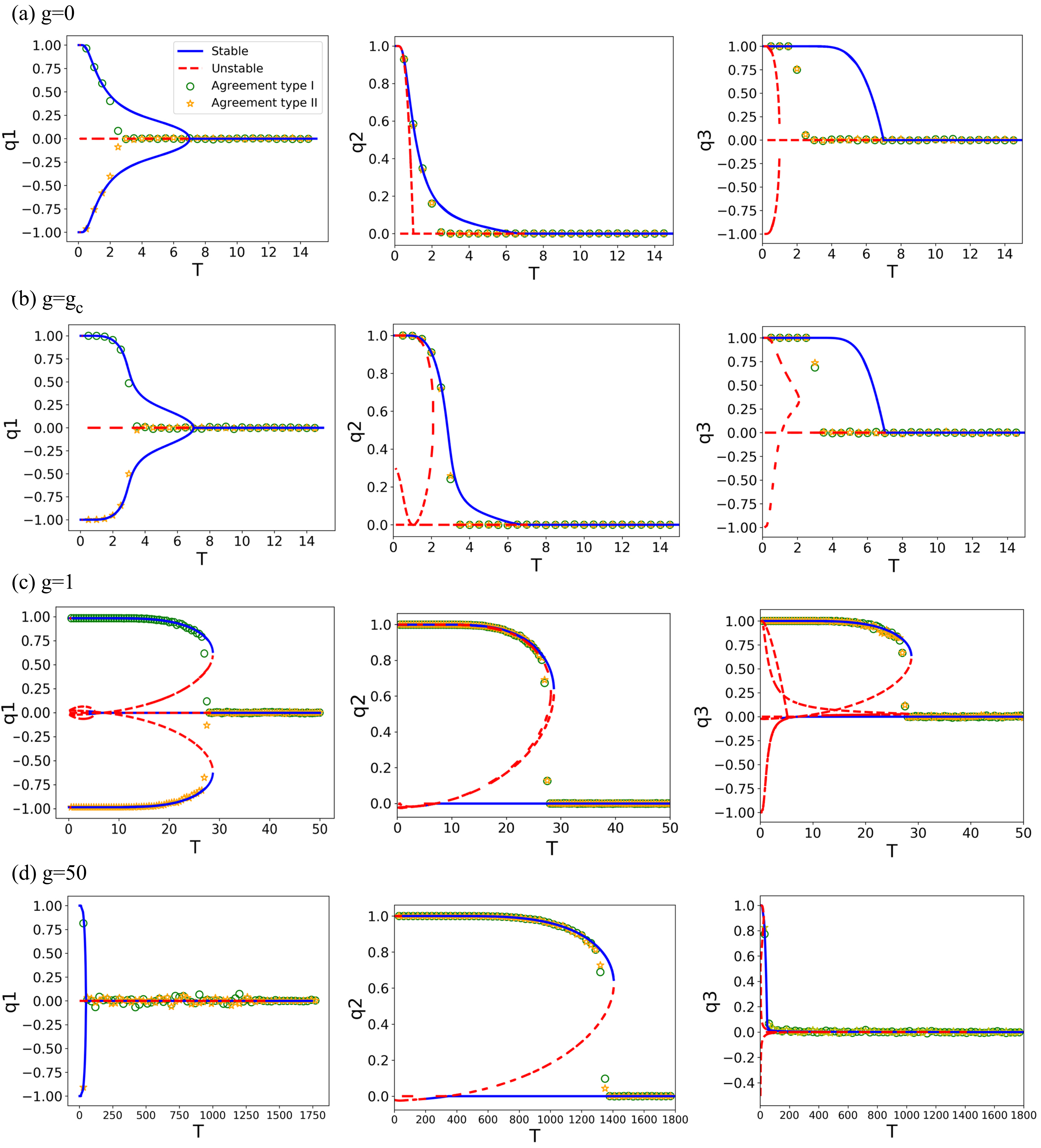}}~~~~
	\end{center}
	\caption{Two-point correlation functions $q_1$, $q_2$ and $q_3$ versus temperature $T$ for different values of $g$. We have a fully connected network of size $n=50$. The solid (blue) and dashed (red) lines show stable and unstable fixed points respectively. Below the tricritical point $g_{c}$ at $g=0$, the phase transition is continuous and there is one critical temperature (a). At $g_c$, the type of transitions changes (b), and above this point, at $g=1$ and $g=50$, there is a discontinuous phase transition and we have a cold and a hot critical temperature (the lower critical temperature is named as cold and higher critical temperature is named as hot)(c, d). At $g=50$ that the node's contribution is significantly small, the parameters $q_1$ and $q_3$ become meaningless. Symbols show numerical simulation results for two different initial configurations, which are in agreement with the mean-field approach (solid lines).
	}
	\label{fig:big-fig}
\end{figure*}

In general, the average value of a variable $x$ is defined as $\langle x\rangle=\sum_{G} x\,P(G)$.	

In order to calculate $m=\langle s_{i}\rangle$, we will extract all the terms including an specific node $s_{i}$ from the Hamiltonian. This part of Hamiltonian is written as the following:

\begin{equation}\label{eq3}
-\mathcal{H}_{i}=s_{i}{\sum_{i\neq l} \sigma_{il}s_{l}}.
\end{equation}

Hence we can rewrite the Hamiltonian as $ \mathcal{H}=\mathcal{H}_1+ \mathcal{H}_{i} $, in which $\mathcal{H}_{i}$ is sum of all terms containing $s_i$ and the $\mathcal{H}_1$ indicates all the remaining terms. With the above description, the average state of a node is obtained by,

\begin{equation}\label{eq8}
\begin{aligned}
m&=\frac{\sum_{G_1}{s_{i}e^{-\beta\mathcal{H}(G_1)}}}{\sum_{G_1} e^{-\beta\mathcal{H}(G_1)}}=\frac{\sum_{s\neq s_{i}}{e^{-\beta\mathcal{H}_1}}\sum_{s=s_{i}}{s_{i}e^{-\beta\mathcal{H}_{i}}}}{\sum_{s\neq s_{i}}{e^{-\beta\mathcal{H}_1}}\sum_{s=s_i}{e^{-\beta\mathcal{H}_i}}}\\
&=\frac{\langle{e^{-\beta\mathcal{H}_{i}(s_{i}=+1)}-e^{-\beta\mathcal{H}_{i}(s_{i}=-1)}}\rangle_{G_1}}{\langle{e^{-\beta\mathcal{H}_{i}(s_{i}=+1)}+e^{-\beta\mathcal{H}_{i}(s_{i}=-1)}}\rangle_{G_1}}.
\end{aligned}
\end{equation}

where $G_1$ is the all network configurations in which the term $s_{i}$ is not included. We can apply the mean field approximation and substitute $s_{i}\sigma_{ij}$ with $q_1$. Hence we will have

\begin{equation}\label{eq9}
m=\frac{e^{\beta(n-1)q_{1}}-e^{-\beta(n-1)q_{1}}}{e^{\beta(n-1)q_{1}}+e^{-\beta(n-1)q_{1}}}	=\tanh[\beta(n-1)q_{1}].	
\end{equation}

Similarly we can calculate the average value of a link, $p=\langle\sigma_{ij}\rangle$, by rewriting the Hamiltonian as $ \mathcal{H}=\mathcal{H}_{ij}+\mathcal{H}_2 $, in which $\mathcal{H}_{ij}$ implies the all terms in the Hamiltonian containing the link between $i$ and $j$,

\begin{equation}\label{eq10}
-\mathcal{H}_{ij}=\sigma_{ij}\Big(s_i\,s_j+ g\,{\sum_{k\neq i,j}{\sigma_{jk}\sigma_{ki}}}\Big).
\end{equation}

and the $\mathcal{H}_2$ counts the remaining terms. Hence, the mean value of $p$ can be written as the following:

\begin{equation}\label{eqp2}
	\begin{aligned}
		p&=\frac{\sum_{G_2}{\sigma_{ij}e^{-\beta\mathcal{H}(G_2)}}}{\sum_{G_2} e^{-\beta\mathcal{H}(G_2)}}=\frac{\sum_{\sigma\neq \sigma_{ij}}{e^{-\beta\mathcal{H}_2}}\sum_{\sigma=\sigma_{ij}}{s_{i}e^{-\beta\mathcal{H}_{ij}}}}{\sum_{\sigma\neq \sigma_{ij}}{e^{-\beta\mathcal{H}_2}}\sum_{\sigma=\sigma_{ij}}{e^{-\beta\mathcal{H}_{ij}}}}\\
		&=\frac{\langle{e^{-\beta\mathcal{H}_{ij}(\sigma_{ij}=+1)}-e^{-\beta\mathcal{H}_{ij}(\sigma_{ij}=-1)}}\rangle_{G_2}}{\langle{e^{-\beta\mathcal{H}_{ij}(\sigma_{ij}=+1)}+e^{-\beta\mathcal{H}_{ij}(\sigma_{ij}=-1)}}\rangle_{G_2}}.
	\end{aligned}
\end{equation}

where $G_2$ is the all network configurations in which in $\sigma_{ij}$ is not included.
Using the mean field approximation ($\sigma_{ij}\sigma_{jk}=q_2$, $s_is_j=q_3$) we obtain:

\begin{equation}\label{eq12}
\begin{aligned}
	p&=\frac{e^{\beta(q_{3}+g(n-2)q_{2})}-e^{-\beta(q_{3}+g(n-2)q_{2})}}{e^{\beta(q_{3}+g(n-2)q_{2})}+e^{-\beta(q_{3}+g(n-2)q_{2})}}\\
	&=\tanh [\beta\,(q_{3}+g\,(n-2)\,q_{2})].
\end{aligned}
\end{equation}


Using the same method, we are able to calculate the two-points correlations $q_1$, $q_2$ and $q_3$. 
To derive $q_1$, we separate all terms that contain $ s_i $, $ \sigma_{ij} $ and $s_i\sigma_{ij} $. Therefore, it is obtained:
\begin{equation}\label{calc1Eq1}
\mathcal{H}=\mathcal{H}_3+ \mathcal{H}_{_{\tiny\nodelink}}.
\end{equation}	
in which $\mathcal{H}_{_{\tiny\nodelink}}$ indicates all the terms that contain $ s_i $, $ \sigma_{ij} $ and $s_i\sigma_{ij} $, and $\mathcal{H}_3$ is the remaining terms. We can wrire $\mathcal{H}_{_{\tiny\nodelink}}$ as the following:

\begin{equation}\label{calc1Eq2}
-\mathcal{H}_{_{\tiny\nodelink}}=s_i \sum_{\ell\neq i,j}\sigma_{i\ell}\,s_\ell+g\sigma_{ij}\sum_{\ell\neq i,j}\sigma_{j\ell}\sigma_{\ell i}+ s_i\,\sigma_{ij}\,s_j.
\end{equation}

Therefore we have:
\begin{widetext}
\begin{equation}\label{calc1Eq3}
\begin{aligned}
	&q_1=\frac{\sum_{\{\sigma\neq\sigma_{ij},\, s\neq s_i\}}e^{-\beta\mathcal{H}_3}\sum_{\sigma_{ij}=\pm 1,\,s_i=\pm 1}s_i\,\sigma_{ij}\,e^{-\beta\mathcal{H}_{_{\tiny\nodelink}}}}{\sum_{\{\sigma\neq\sigma_{ij},\, s\neq s_i\}}e^{-\beta\mathcal{H}_3}\sum_{\sigma_{ij}=\pm 1,\,s_i=\pm 1}\,e^{-\beta\mathcal{H}_{_{\tiny\nodelink}}}}\\
	&=\frac{\left\langle e^{-\beta\mathcal{H}_{_{\tiny\nodelink}}(s_i=1,\sigma_{ij}=1)}-e^{-\beta\mathcal{H}_{_{\tiny\nodelink}}(s_i=-1,\sigma_{ij}=1)}-e^{-\beta\mathcal{H}_{_{\tiny\nodelink}}(s_i=1,\sigma_{ij}=-1)}+e^{-\beta\mathcal{H}_{_{\tiny\nodelink}}(s_i=-1,\sigma_{ij}=-1)}\right\rangle_{G_3}}{ \left\langle e^{-\beta\mathcal{H}_{_{\tiny\nodelink}}(s_i=1,\sigma_{ij}=1)}+e^{-\beta\mathcal{H}_{_{\tiny\nodelink}}(s_i=-1,\sigma_{ij}=1)}+e^{-\beta\mathcal{H}_{_{\tiny\nodelink}}(s_i=1,\sigma_{ij}=-1)}+e^{-\beta\mathcal{H}_{_{\tiny\nodelink}}(s_i=-1,\sigma_{ij}=-1)}\right\rangle_{G_3}},\\
	&=\frac{e^{-\beta(n-2)(q_{1}+gq_{2})+\beta m}+e^{\beta(n-2)(q_{1}+gq_{2})+\beta m}-e^{-\beta(n-2)(q_{1}-gq_{2})-\beta m}-e^{\beta(n-2)(q_{1}-gq_{2})-\beta m}
	}{e^{-\beta(n-2)(q_{1}+gq_{2})+\beta m}+e^{\beta(n-2)(q_{1}+gq_{2})+\beta m}+e^{-\beta(n-2)(q_{1}-gq_{2})-\beta m}+e^{\beta(n-2)(q_{1}-gq_{2})-\beta m}}\\
&\equiv f_1(q_1,\,q_2\,;\; \beta ,\, n,\, g).
	\\
\end{aligned}
\end{equation}
\end{widetext}

\begin{figure}[!t]
	\centering
    \includegraphics[width=\linewidth]{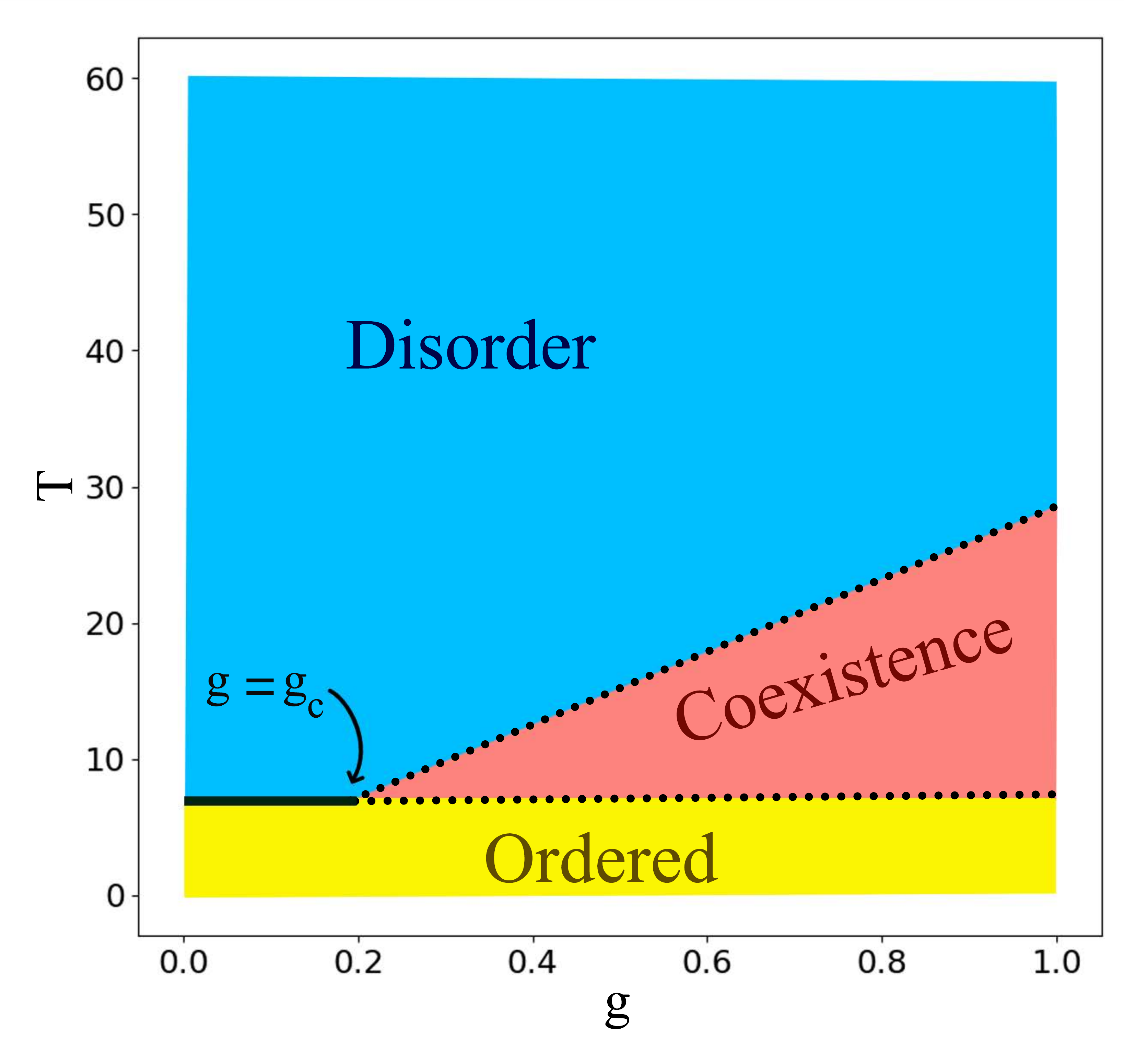}
	\caption{ Phase diagram of a fully-connected network with $n=50$ in the space $(T, g)$. The phases with different number of stable fixed points are specified with different colors. The model shows one, two or three fixed points in the disorder phase (blue), ordered (yellow) and coexistence (pink) areas respectively. Below the tricritical point, the solid line shows there is only one critical temperature where the phase transition is continuous. The two dotted lines which restrict the coexistence area is in respect of cold and hot critical temperatures where there is a discontinuous phase transition.
	}
	\label{fig:fig2}
	
\end{figure}


Similarly we can separate the terms of Hamiltonian which contain links  $ \sigma_{ij} $ and $ \sigma_{jk} $ and show theses terms with $ \mathcal{H}_{\vee} $. The remaining terms of Hamiltonian is represented by $ \mathcal{H}_4 $. Hence we rewrite Hamiltonian $\mathcal{H}$ as follows:
\begin{equation}\label{calc2Eq1}
\mathcal{H}=\mathcal{H}_4+ \mathcal{H}_{\vee}
\end{equation}	
where $ \mathcal{H}_{\vee} $ is written as

\begin{equation}\label{calc2Eq2}
\begin{aligned}
-\mathcal{H}_{\vee}&=\sigma_{ij}\Big(s_i\,s_j+g\sum_{\ell\neq i,j,k}\sigma_{j\ell}\sigma_{\ell i}\Big)\\
&+ \sigma_{jk}\Big(s_j\,s_k+g\sum_{\ell\neq i,j,k}\sigma_{j\ell}\sigma_{\ell i}\Big)+g\;\sigma_{ij}\sigma_{jk}\sigma_{ki}.\\
\end{aligned}
\end{equation}	

Consequently, the link-link correlation $ q_2 $ is obtained as:

\begin{widetext}
\begin{equation}\label{calc2Eq3}
\begin{aligned}
q_{2}&=\frac{\sum_{\{\sigma\neq\sigma_{ij},\, \sigma\neq \sigma_{jk}\}}e^{-\beta\mathcal{H}_4}\sum_{\sigma_{ij}=\pm 1,\,\sigma_{jk}=\pm 1}\sigma_{ij}\,\sigma_{jk}\,e^{-\beta\mathcal{H}_{_{\vee}}}}{\sum_{\{\sigma\neq\sigma_{ij},\, \sigma\neq \sigma_{jk}\}}e^{-\beta\mathcal{H}_4}\sum_{\sigma_{ij}=\pm 1,\,\sigma_{jk}=\pm 1}\,e^{-\beta\mathcal{H}_{_{\vee}}}}\\
&=\frac{\left\langle e^{-\beta\mathcal{H}_{_{\vee}}(\sigma_{ij}=1,\sigma_{jk}=1)}-e^{-\beta\mathcal{H}_{_{\vee}}(\sigma_{ij}=-1,\sigma_{jk}=1)}-e^{-\beta\mathcal{H}_{_{\vee}}(\sigma_{ij}=1,\sigma_{jk}=-1)}+e^{-\beta\mathcal{H}_{_{\vee}}(\sigma_{ij}=-1,\sigma_{jk}=-1)}\right\rangle_{G_4}}{ \left\langle e^{-\beta\mathcal{H}_{_{\vee}}(\sigma_{ij}=1,\sigma_{jk}=1)}+e^{-\beta\mathcal{H}_{_{\vee}}(\sigma_{ij}=-1,\sigma_{jk}=1)}+e^{-\beta\mathcal{H}_{_{\vee}}(\sigma_{ij}=1,\sigma_{jk}=-1)}+e^{-\beta\mathcal{H}_{_{\vee}}(\sigma_{ij}=-1,\sigma_{jk}=-1)}\right\rangle_{G_4}},\\
&=\frac{e^{4\beta g(n-3)q_2 + 4\beta q_3} - 2\,e^{2\beta g(n-3)q_2 + 2\beta q_3 - 2\beta g p} + 1}{e^{4\beta g(n-3)q_2 + 4\beta q_3} + 2\,e^{2\beta g(n-3)q_2 + 2\beta q_3 - 2\beta g p} + 1}\\
\\
&\equiv f_2(q_2,\,q_3\,;\; \beta ,\, n,\, g).\\
\end{aligned}
\end{equation}	
\end{widetext}

Finally we reach the node-node correlation $q_3$, by separating the terms which contain the nodes $ s_i $ and $ s_j $. This part of Hamiltonian is represented by $ \mathcal{H}_{_{\circ\circ}} $:
\begin{equation}\label{calc3Eq2}
\begin{aligned}
-\mathcal{H}_{_{\circ\circ}}&=s_i \sum_{\ell\neq i,j}\sigma_{i\ell}\,s_\ell+s_j \sum_{\ell\neq i,j}\sigma_{j\ell}\,s_\ell+s_i\, \sigma_{ij}\,s_j.\\
\end{aligned}
\end{equation}	

The remaining terms of Hamiltonian is given by $\mathcal{H}_5$, such that the Hamiltonian is written as the following:

\begin{equation}\label{calc3Eq1}
\mathcal{H}=\mathcal{H}_5+ \mathcal{H}_{_{\circ\circ}}.
\end{equation}	

Therefore, we can find $q_3$ as follows:
\begin{widetext}
\begin{equation}\label{calc3Eq3}
\begin{aligned}
q_3 &=\frac{\sum_{\{s\neq s_i,\, s\neq s_j\}}e^{-\beta\mathcal{H}_5}\sum_{s_i=\pm 1,\,s_j=\pm 1}s_i\,s_j\,e^{-\beta\mathcal{H}_{_{\circ\circ}}}}{\sum_{\{s\neq s_i,\, s\neq s_j\}}e^{-\beta\mathcal{H}_5}\sum_{s_i=\pm 1,\,s_j=\pm 1}\,e^{-\beta\mathcal{H}_{_{\circ\circ}}}}\\
&=\frac{\left\langle e^{-\beta\mathcal{H}_{_{\circ\circ}}(s_i=1,s_j=1)}-e^{-\beta\mathcal{H}_{_{\circ\circ}}(s_i=-1,s_j=1)}-e^{-\beta\mathcal{H}_{_{\circ\circ}}(s_i=1,s_j=-1)}+e^{-\beta\mathcal{H}_{_{\circ\circ}}(s_i=-1,s_j=-1)}\right\rangle_{G_5}}{ \left\langle e^{-\beta\mathcal{H}_{_{\circ\circ}}(s_i=1,s_j=1)}+e^{-\beta\mathcal{H}_{_{\circ\circ}}(s_i=-1,s_j=1)}+e^{-\beta\mathcal{H}_{_{\circ\circ}}(s_i=1,s_j=-1)}+e^{-\beta\mathcal{H}_{_{\circ\circ}}(s_i=-1,s_j=-1)}\right\rangle_{G_5}},\\
&=\frac{e^{2\beta (n-2)q_1 + 2\beta p} + e^{- 2\beta (n-2)q_1 + 2\beta p} - 2}{e^{2\beta (n-2)q_1 + 2\beta p} + e^{- 2\beta (n-2)q_1 + 2\beta p} + 2}\\
\\
&\equiv f_3(q_1,\,q_2,\,q_3\,;\; \beta ,\, n,\, g).\\
\end{aligned}
\end{equation}	
\end{widetext}

\begin{figure*}[t]
\begin{center}
\scalebox{0.35}{\includegraphics[angle=0]{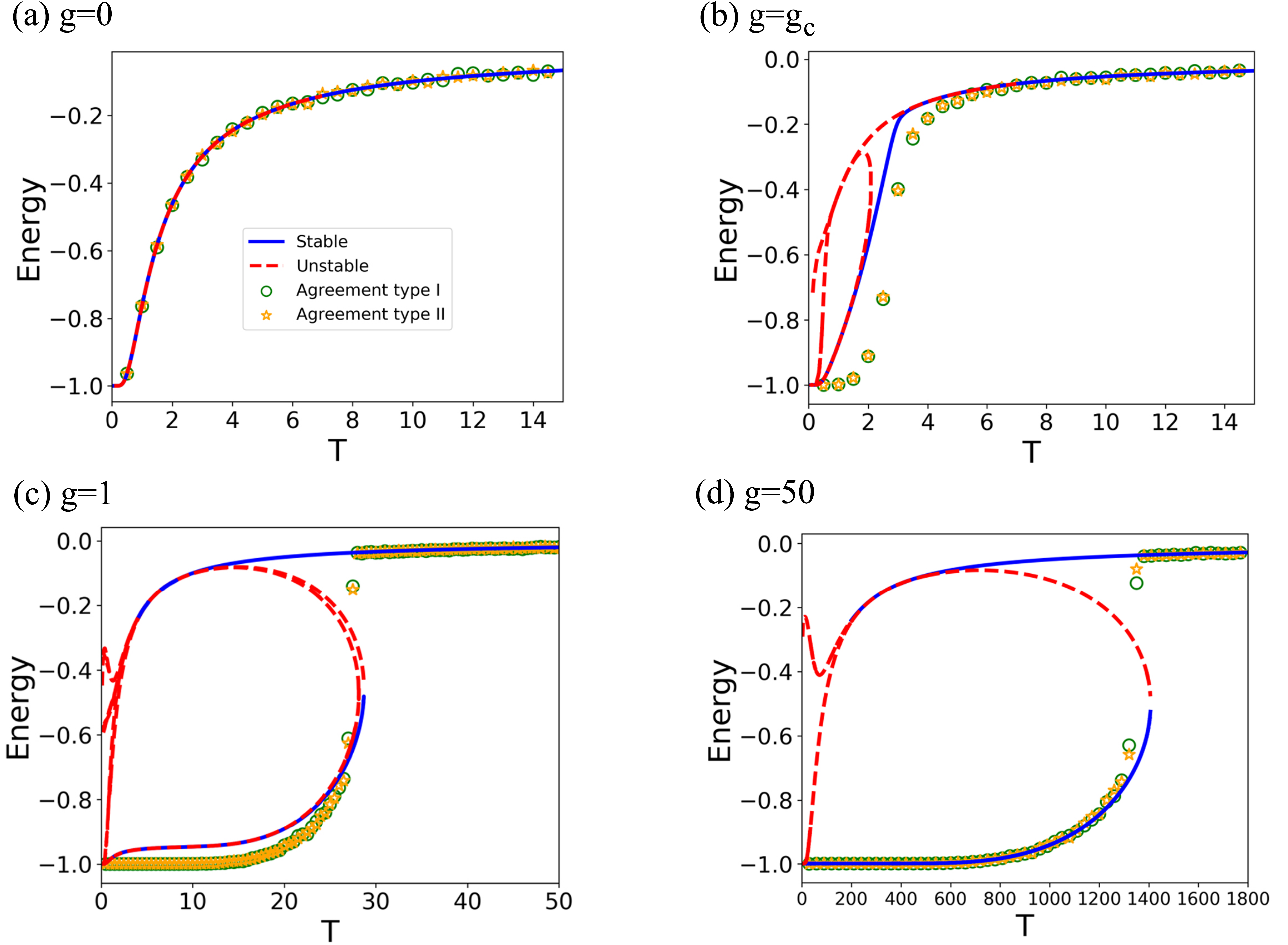}}~~~~
\end{center}
\caption{Total energy of a fully-connected network of size  $n=50$ versus temperature $T$ and for different values of $g$. The solid (blue) and dashed (red) lines show stable and unstable fixed points respectively. At $g=0$, the system shows a continuous phase transition and there is one critical temperature. For $g=1$ and $g=50$, a discontinuous transition emerges where there are a cold and a hot critical temperature point. Symbols show numerical simulation results for two different initial configurations, which are in agreement with the mean-field approach (solid lines).   
}
\label{fig:big-fig5}
\end{figure*}

There is a set of self-consistence equations as the following:
\begin{equation}\label{eq13}
\begin{aligned}
q_1&= f_1(q_1,\,q_2\,;\; \beta ,\, n,\, g),	\\
q_2&= f_2(q_2,\,q_3\,;\; \beta ,\, n,\, g),\\
q_3&= f_3(q_1,\,q_2,\,q_3\,;\; \beta ,\, n,\, g).
\end{aligned}
\end{equation}
\begin{figure*}[t]
\begin{center}
\scalebox{0.38}{\includegraphics[angle=0]{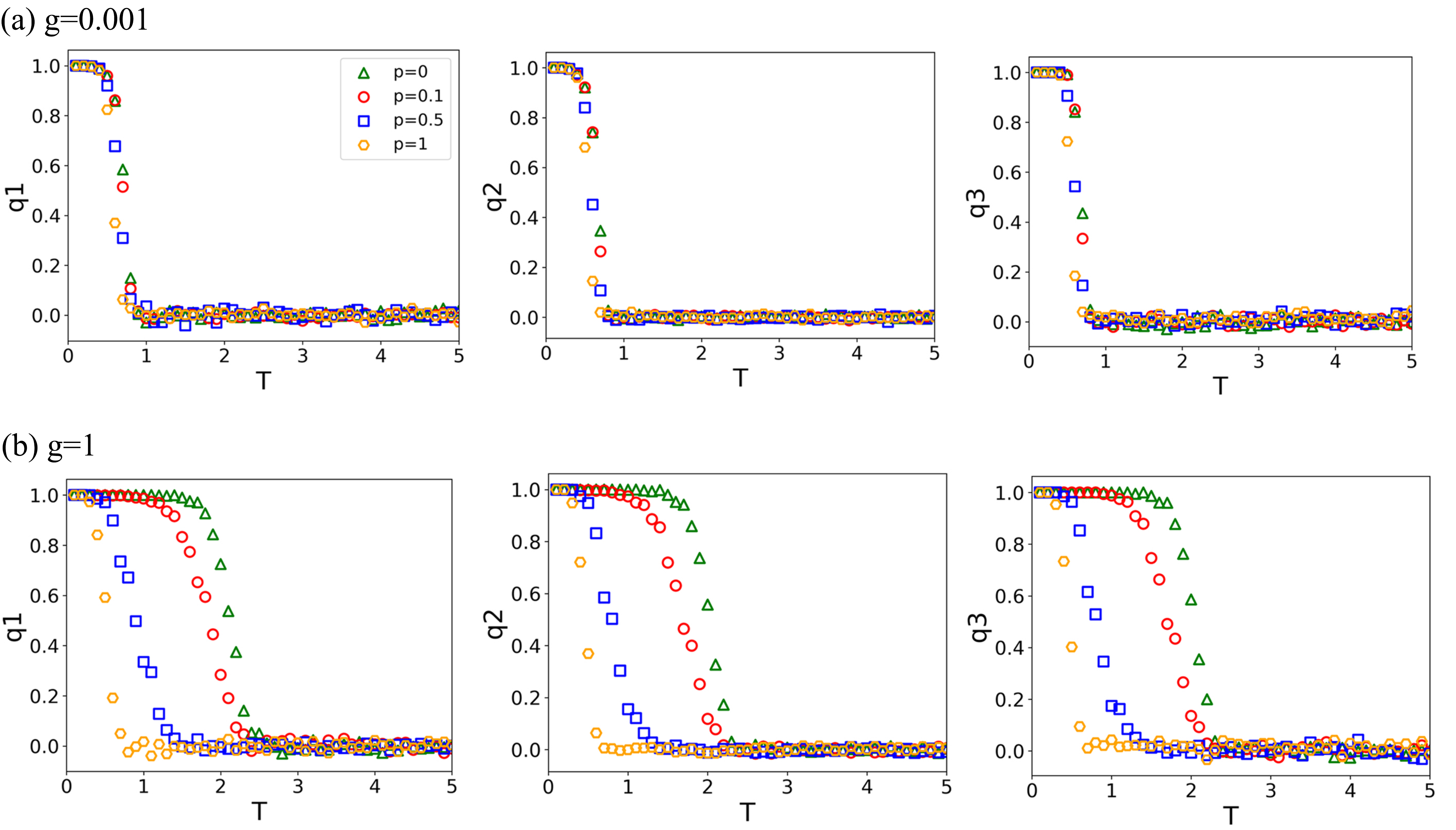}}~~~~
\end{center}
\caption{Two-point correlations $q_1$, $q_2$ and $g_3$ versus temperature $T$ on small-world networks of size $n=50$ and mean degree $\langle k \rangle=8$. Symbols show the numerical simulation results for different rewiring probabilities $p$. In $(a)$ $g=0.001$, the network randomness has no effect on the correlation behavior. In $(b) $ $g=1$, and critical temperature $T_c$ shifts to the higher values for more ordered networks.}
\label{fig:big-fig2}
\end{figure*}

Solving Eqs.~(\ref{eq13}) numerically, we obtain all possible solutions as a set of  $ (q_1^* , q_2^*, q_3^*) $. The stability condition for solutions can be checked by calculating the eigenvalues of Jacobian matrix of Eqs.~(\ref{eq13}) (see Appendix \ref{appendixI}).

The stable and unstable fixed points are plotted as a function of $T$ in Fig.~\ref{fig:big-fig}. At $g=0$, the two-point correlation functions $q_1$, $q_2$ and $g_3$ show a second order phase transition, whereas at $g=1$ the first order phase transition appears. It means that with increasing $g$, the phase transition type changes from second to first order at a tricritical point $g_c$.

In a network of size $n$, the number of node-link-node interactions is $\binom{n}{2}$ and there is $\binom{n}{3}$ of triangles. On this account, we can estimate the tricritical point $g_c$ as follows,
\begin{equation}\label{gc}
	g_{c}\simeq\frac{\binom{n}{2}}{\binom{n}{3}}=\frac{3}{(n-2)}.
\end{equation}
Therefore, we expect the type of phase transition around this point switches from continuous to discontinuous. This is clearly shown in Fig.~\ref{fig:big-fig} where the order parameters of the network are plotted at different values of $g$. Depending on $g$, two types of phase transitions are observed, the continuous type, at $g=0$ with one critical point and the discontinuous type, at $g=1$ and $g=50$, where we can see that there are cold and hot critical temperatures. Actually, the cold critical temperature is assigned to the point at which the two stable fixed points existing at low temperatures, convert to three stable fixed points. The hot critical temperature is the point that three stable fixed points change to one stable fixed point.

\begin{figure*}[t]
	\begin{center}
		\scalebox{0.38}{\includegraphics[angle=0]{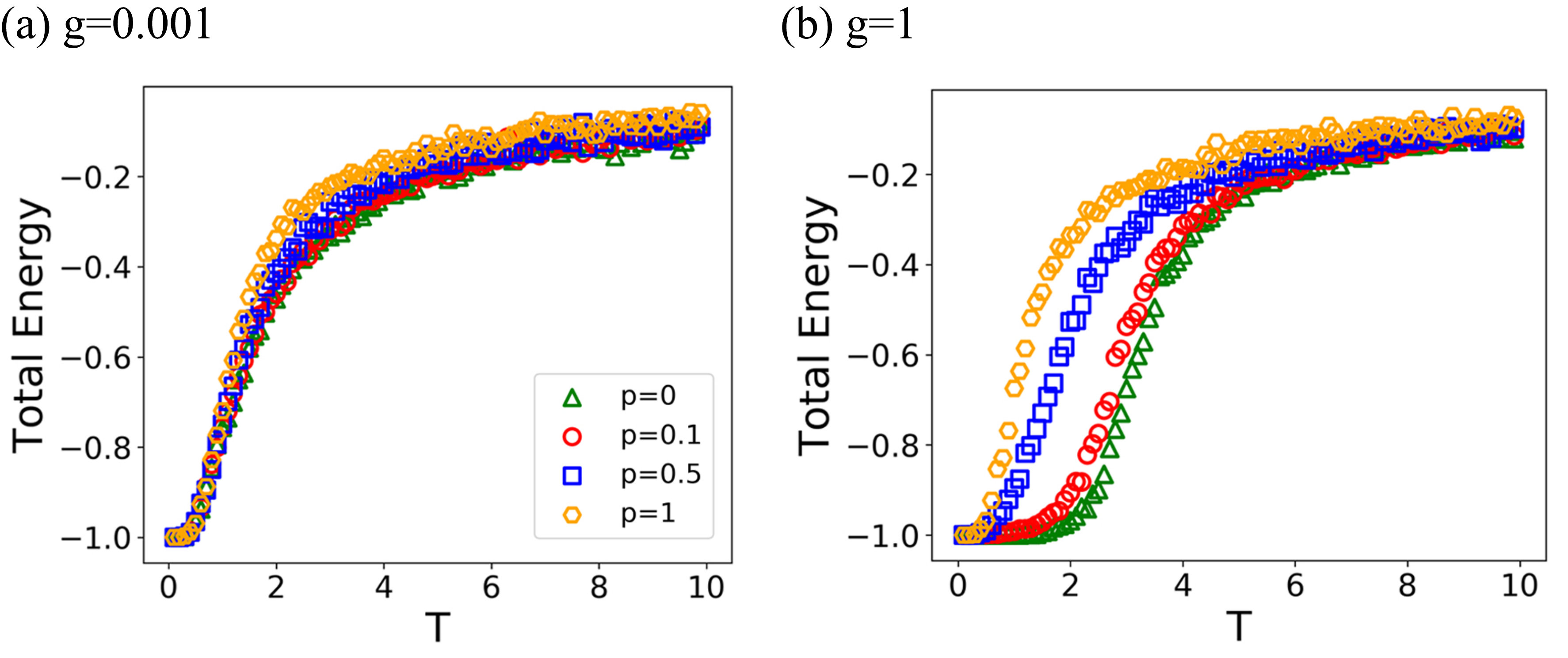}}~~~~
	\end{center}
	\caption{Total energy of small-world networks of size $n=50$ and mean degree $\langle k \rangle=8$ versus temperature $T$ and for different values of $g$. Symbols show the numerical simulation results for different rewiring probabilities $p$. In $(a)$ $g=0.001$, the network randomness has no effect on the energy behavior. In $(b)$ $g=1$, and critical temperature $T_c$ shifts to the higher values for more ordered networks.}
	\label{fig:big-fig23}
\end{figure*}

The number of stable fixed points is demonstrated in Fig.~\ref{fig:fig2} in the plane $(T,g)$. As illustrated in this figure, depending on the changes in the number of stable fixed points at different values of $g$ and $T$, we can detect three areas distinguished by different colors. At low temperatures, independent of the value of $g$, there are two stable fixed points $ (q_1^*=\pm 1,\,q_2^*=1,\,q_3^*=1) $. These solutions refer to situations in which all members of society agree with each other and all relationships between them are friendly. In this case, the system is at its minimum overall tension or energy. This area is named as ordered and has a yellow color in Fig. ~\ref{fig:fig2}.
With increasing $T$, below the tricritical point, a new fixed point $ (q_1^*=0,\,q_2^*=0 ,\,q_3^*=0) $ emerges. This fixed point shows a  disorder phase in which all correlations are zero and it is the only solution for temperatures greater than the critical temperature $T_c$. The disorder  phase area is colored by blue in Fig.~\ref{fig:fig2}.

When $g$ is increased, above the tricritical point, the term of the structural balance becomes more important against the coevolutionary balance. In this case, there are three stable solutions and both of the balance terms have the chance to occur. We call this area coexistence (pink area). Among these three fixed points, for $q_1$ two fixed points are symmetric and positive and negative correlations can exist. $q_2$ and $q_3$ show a non-zero positive solution and a trivial fixed point $q_2^*=q_3^*=0$. 
As the temperature increases, after passing the hot critical temperature the system will again get into the disorder phase area.
We can also see that with increasing $g$, the hot temperature critical points shift to the higher values. 

In Fig.~\ref{fig:fig2} we have shown the tricrital point, obtained through the mean field approximation, where the coexistence appears and the two regimes of second and first order phase transitions switches. The critical temperature at continuous phase transition regime are shown with solid black line and at the discontinuous phase transition regime with black dotted lines.
Moreover, in comparison of two balance types, at high enough amounts of $g$ that structural balance is dominant, the critical temperature occurs at higher temperatures rather than the small amounts of $g$ where the co-evolutionary balance plays more effective role.

Besides, we find the total energy using mean-field approximation. For this purpose, we calculate average of each term of the Hamiltonian (Eq.~(\ref{eq1})).
In order to obtain the average of first term of the Hamiltonian related to the contribution of the node-link-node triplets which we represent it with notation $E_{_{\tiny\circ\multimap}}$, we have considered all the terms in the Hamiltonian containing node-link-node triplets as below:

\begin{equation}\label{calc4Eqa1}
	\begin{aligned}
		-\mathcal{H}_{\nodeLinkNode}=s_i \sum_{\ell\neq i,j}\sigma_{i\ell}\,s_\ell+s_j \sum_{\ell\neq i,j}\sigma_{j\ell}\,s_\ell+ s_i\,\sigma_{ij}\,s_j.
	\end{aligned}
\end{equation}

For the average of second term of the Hamiltonian, in respect of the link-link-link triplets (triangles) contribution, represented by $ E_{_{\triangle}} $ in the following, we will separate all the terms in the Hamiltonian including link-link-link interactions. Therefore we write:

\begin{equation}\label{calc4Eqa2}
\begin{aligned}
-\mathcal{H}_{_{\triangle}}&=g\Big(\sigma_{ij}\sum_{\ell\neq i,j,k}\sigma_{i\ell}\sigma_{\ell j}+\sigma_{jk}\sum_{\ell\neq i,j,k}\sigma_{j\ell}\sigma_{\ell k}\\
&+\sigma_{ki}\sum_{\ell\neq i,j,k}\sigma_{k\ell}\sigma_{\ell i}+\sigma_{ij}\sigma_{jk}\sigma_{ki}\Big).
\end{aligned}
\end{equation}

We approximate equations Eq. (\ref{calc4Eqa1}) and Eq. (\ref{calc4Eqa2}) with the mean field method to reach the above-mentioned averages for $E_{_{\tiny\circ\multimap}}$ and $ E_{_{\triangle}} $.
Accordingly, we have $E_{_{\tiny\circ\multimap}}$ as: 
\begin{equation}\label{avNL}
	\begin{aligned}
		E_{_{\tiny\circ\multimap}}&=\frac{-1+e^{2\beta}}{1+e^{2\beta}}\\
	\end{aligned}
\end{equation}
 and also $ E_{_{\triangle}} $ is calculated as:
 \begin{equation}\label{avLL}
 	\begin{aligned}
 		 E_{_{\triangle}} &=\frac{-1-3e^{4g\beta(n-3)q_2}+3e^{2g\beta(1+(n-3)q_2)}+e^{2g\beta(1+3(n-3)q_2)}}{+1+3e^{4g\beta(n-3)q_2}+3e^{2g\beta(1+(n-3)q_2)}+e^{2g\beta(1+3(n-3)q_2)}}.\\
 	\end{aligned}
 \end{equation}
The normalized energy is
\begin{equation}\label{calc4Eqa3}
\begin{aligned}
E&=-\frac{n_{\nodeLinkNode}E_{\nodeLinkNode}+n_{_{\triangle}}E_{_{\triangle}}}{n_{\nodeLinkNode}+g\,n_{_{\triangle}}}
\end{aligned}
\end{equation}
where $ n_{\nodeLinkNode} $ is the number of node-link-node triplets and $ n_{_{\triangle}} $ is the number of triangles. In a fully connected network these numbers are $ n_{\nodeLinkNode}=\binom{n}{2} $ and  $ n_{_{\triangle}}=\binom{n}{3} $.

Therefore, by substituting Eq.~(\ref{avNL}) and Eq.~(\ref{avLL}) in Eq.~(\ref{calc4Eqa3}) the total energy will be obtained as:
\begin{widetext}
 \begin{equation}\label{tote}
	\begin{aligned}
		E &=\frac{\frac{3(-1+e^{2\beta})}{1+e^{2\beta}}+\frac{g(n-2)(-1-3e^{4g\beta(n-3)q_2}+3e^{2g\beta(1+(n-3)q_2)}+e^{2g\beta(1+3(n-3)q_2)})}{1+3e^{4g\beta(n-3)q_2}+3e^{2g\beta(1+(n-3)q_2)}+e^{2g\beta(1+3(n-3)q_2)}}}{3+g(n-2)}\\
	\end{aligned}
\end{equation}
\end{widetext}

Fig.~\ref{fig:big-fig5} shows the behavior of the total energy as a function of temperature. It is clear in this figure that the type of phase transition changes depending on the value of $g$. In the other words, when the structural balance has more power and the value of $g$ is high enough, the phase transition is discontinuous, however at low $g$ values, below the triciritical point the phase transition is continuous.

\section{Simulation}\label{simulation}
	
To confirm the results we can perform numerical simulations. First we consider a fully-connected network with $n=50$. With an initial condition for the state of nodes and links, we calculate Hamiltonian $\mathcal{H}$ from Eq.~(\ref{eq1}). We use the Metropolis algorithm and update the state of the nodes and links. At each time step with probability $ p $ a random node and with probability $ (1-p)$ a random link is chosen and flipped its state. The probability $p=\frac{n}{n+n(n-1)/2}$, is the ratio of the number of the nodes to the total number of nodes and links. In the new configuration Hamiltonian $\mathcal{H}$ is again calculated and if it decreases, the flip is accepted. Otherwise with the probability $e^{-(\mathcal{H}_{t+1}-\mathcal{H}_{t})/T}$ we keep the flip. The process continues until the system reaches a stationary state.

In Fig.~\ref{fig:big-fig} we compare the mean field solutions, obtained in the previous section, with the simulation results for different values of $g$. We consider two types of initial conditions: 1) all the links and nodes are positive (agreement type I) and 2) the links are all positive and the nodes are all negative (agreement type II).  The results show a good correspondence, however for the low values of $g$ $(g<g_c)$, the simulation dose not properly confirm the mean field approximation solutions. With increasing $g$ and reinforcing the contribution of structural balance term, the agreement between the mean field approximation and the numerical simulations is well observed (Figs.~2 (c)-(d)).

In the following we consider the model on small world networks. Let us start with a ring having $n=50$ nodes and node degree $k=8$. Moving clockwise, for every node we select randomly a link that connects that node to one of its neighbors, and rewire it with probability $p$. We continue this process until each link in the original ring has been considered once. The parameter $p$ measures the randomness of the resulting network. For $p = 0$ the network is regular and there are many triangles (high clustering coefficient), while for $p = 1$ all links are rewired and the resulting network is a random network with low clustering coefficient. Hence for the small values of $g$, the degree of network's randomness does not have any effect. As we can see in Fig.~\ref{fig:big-fig2} the behavior of correlation functions $q_1$, $q_2$, $q_3$ is independent on $p$. However with increasing $g$, the interaction of triangles find more contribution and the structural balance become more effective. Thus the correlation functions behave differently for the different values of rewiring probability $p$. As we can see, the transition points shift to the smaller temperature with increasing the network randomness. Similar behavior is observed in the total energy of the small-world networks (Fig.~\ref{fig:big-fig23}). 

\section{Conclusion}\label{Conclusion}

 In this work we have considered the structural Heider balance in which the interactions have a global effect, as each link contributes to so many triangles in the network structure and the coevolutionary balance in which the links only connect the local nodes. Moreover, studies show structural balance represents a discontinuous phase transition while the coevolutionary balance results in a continuous phase transition. Given the fact that one can not ignore the existence of each term in any arbitrary system, we have simultaneously considered both terms in order to evaluate the role of each balance type in the final balance state of the system. The factor $g$ in the Hamiltonian of the model controls the contribution of the structural versus the coevolutionary balance. Using statistical mechanics methods and the mean field approximation, we presented analytical relations of the two point correlations as well as the total energy of a fully-connected network and compared the results with the simulations. Finally, we conclude our investigations in this study with the following points:

	- We found that in this competition, on equal conditions, when factor $g$ is equal to $1$, the structural balance term is the winner term and we can clearly observe the discontinuous phase transition in the system meaning that the network is propelled by the triad interactions into a global balance and there is more interest in fulfilling the total overall benefits of the system rather than the local benefits of the system agents.

	- At low values of $g$, the coevolutionary term, expressing the local balance, will get more power versus the Heider term and thus it is the dominant term and the phase transition is continuous.

	- Due to the existence of two types of phase transitions, we found a tricritical value for $g$, in which the type of transition switches from continuous to discontinuous.

	- Considering that the coevolutionary balance occurs at lower temperatures rather than the Heider balance, we have observed that in the competition between these two terms, at low enough temperatures and for any values of $g$, the system is in an ordered state. However, at any fixed value of $T$, by increasing the factor $g$, which is equivalent to magnifying the role of the structural balance, the competition is in direction to move into a coexistence region where both balance terms are probable.  In a fully connected network the fragmented and consensus phases are not detected. Thurner et al. showed the simulations for an sparse network and they have confirmed at large enough $g$ values, when the structural balance is the dominant term, the society will be in the state that different ideas will be able to coexist. However, when $g$ is small and the coevolutionary balance is predominant, the society will come to a fragmented state in which there are clusters of people who are friendly with each other but they are unfriendly with the people in other clusters. This means that the society does not sustain the co-existence of opposite ideas. 
	
	- The energy function is the normalization of the summation of the two structural and coevolutionary terms. For $g=0$, there exsits only the coevolutionary term and the energy behaviour is due to the local interactions. However, for $g>g_c$ the triangle interactions (structural term) dominate and the coevolutionary term is not effective.
	
	The correlation function $q_2$ is the order parameter related to the link-link interactions, which stands for the Heider balance discipline, whereas $q_3$ and $q_1$ functions show the order parameters of the node-link and node-node interactions, respectively, and stand for the coevolutionary balance discipline. Our observations show that above the tricritical point $g_c$, where the structural balance wins the competition, the parameters $q_1$ and $q_3$ still exist. In other words, the dominance of structural balance pushes the transition temperature of the node-node and node-link order parameters to a higher value (the critical temperature attributed to the structural balance). However for $g<g_c$, the order parameter $q_2$, obeying coevolutionary balance discipline, shows the same transition point as the coevolutionary balance. In particular for $g=0$, where the effect of the structural balance is completely removed, the order parameter $q_2$ becomes zero in a low temperature (the critical temperature attributed to the coevolutionary balance). 

\color{black}{ }
\appendix
\begin{appendices}	
\section{Stability analysis of fixed points}\label{appendixI}

Let us consider fixed point $(q_1^*,\,q_2^*,q_3^*)$. To determine the stability of the fixed point, we consider a nearby solution by an infinitesimal perturbation such that $ (q_1^*+\delta q_1,\,q_2^*+\delta q_2,\,q_3^*+\delta q_3) $. the  fixed point will be mapped by the Eq.~\ref{eq13} to another point in $ (q_1,\,q_2,\,q_3) $ space, as
\begin{equation}\label{eq14}
\begin{aligned}
q_1^*+\delta q_1'&= f_1(q^*_1+\delta q_1,\,q^*_2+\delta q_2\,\,;\;\beta,\,n,\,g),\\
q_2^*+\delta q_2'&= f_2(q^*_2+\delta q_2\,,\,q^*_3+\delta q_3\,;\;\beta,\,n,\,g),\\
q_3^*+\delta q_3'&= f_3(q^*_1+\delta q_1,\,q^*_2+\delta q_2,\,q^*_3+\delta q_3\,;\;\beta,\,n,\,g).
\end{aligned}
\end{equation}
For  $ \delta q_1\ll 1 $, $ \delta q_2\ll 1 $ and $ \delta q_3\ll 1 $, we use the Taylor expansion neglecting the nonlinear terms
\begin{equation}\label{eq15}
\begin{aligned}
q^*_1+\delta q_1'&\approx f_1(q^*_1,\,q^*_2\,;\;\beta,\,n,\,g)
+ \frac{\partial f_1}{\partial q}\Big |_{\substack{(q^*_1,q^*_2)}}\delta q_1  \\
&\qquad\qquad\qquad\qquad\qquad\qquad+ \frac{\partial f_1}{\partial q_2}\Big |_{\substack{(q^*_1,q^*_2)}}\delta q_2,\\
\\
q^*_2+\delta q_2'&\approx f_2(q^*_2,\,q^*_3\,;\;\beta,\,n,\,g)
+ \frac{\partial f_2}{\partial q_2}\Big |_{\substack{(q^*_2,q^*_3)}}\delta q_2  \\
&\qquad\qquad\qquad\qquad\qquad\qquad+ \frac{\partial f_2}{\partial q_3}\Big |_{\substack{(q^*_2,q^*_3)}}\delta q_3,\\
\\
q^*_3+\delta q_3'&\approx f_3(q^*_1,\,q^*_2\,,q^*_3;\;\beta,\,n,\,g)
+ \frac{\partial f_3}{\partial q_1}\Big |_{\substack{(q^*_1,q^*_2,q^*_3)}}\delta q_1  \\
&\qquad\qquad+ \frac{\partial f_3}{\partial q_2}\Big |_{\substack{(q^*_1,q^*_2,q^*_3)}}\delta q_2 + \frac{\partial f_3}{\partial q_3}\Big |_{\substack{(q^*_1,q^*_2,q^*_3)}}\delta q_3. \\
\\
\end{aligned}
\end{equation}
By linearisation we have
\begin{equation}\label{matrixeq}
\left(\begin{array}{c} \delta q_1'\\ \delta q_2' \\ \delta q_3'\end{array}\right)  = \mathbf{J}\left(\begin{array}{c}\delta q_1\\ \delta q_2 \\ \delta q_3\end{array}\right),
\end{equation}
where $\mathbf{J}$ is the Jacobian matrix in the fixed point:
\begin{equation}
\mathbf{J}=
\left(\begin{array}{ccc}
\partial f_1/\partial q_1 & \partial f_1/\partial q_2 & \partial f_1/\partial q_3\\
\partial f_2/\partial q_1 & \partial f_2/\partial q_2 & \partial f_2/\partial q_3\\
\partial f_3/\partial q_1 & \partial f_3/\partial q_2 & \partial f_3/\partial q_3\\ \end{array}\right)_{\substack{(q^*_1,q^*_2,q^*_3)}}.
\end{equation}
By diagonalizing we obtain:
\begin{equation}\label{digonal-form}
\left(\begin{array}{c} {}_{d}\delta q_1' \\ {}_{d}\delta q_2' \\ {}_{d}\delta q_3' \end{array}\right)  =
\left(\begin{array}{ccc}\lambda_1 & 0 & 0\\ 0 & \lambda_2 & 0\\ 0 & 0 & \lambda_3 \end{array}\right)\left(\begin{array}{c} {}_{d}\delta q_1 \\ {}_{d}\delta q_2 \\ {}_{d}\delta q_3 \end{array}\right),
\end{equation}
where vector $ \bm{{}_{d}\delta q'} $ ($ \bm{{}_{d}\delta q} $) is actually the vector $\bm{\delta q'}$ ($\bm{\delta q}$) in diagonal space. If magnitude of all eigenvalues for a fixed point, are smaller than one $ (|\lambda_1|<1, |\lambda_2|<1$ and $|\lambda_3|<1) $ then the fixed point is attractive or stable otherwise it is unstable. This means that the entire vector field around the fixed point is towards this point, and with each iteration we approach the fixed point. In the case where all three eigenvalues are greater than one, the vector field around the fixed point is completely divergent, which means that by each iteration the magnitude of difference vector $ |\bm{{}_{d}\delta q'}| $ will be bigger.

\end{appendices}
%

\end{document}